# Text mining and visualization using VOSviewer

Nees Jan van Eck and Ludo Waltman

Centre for Science and Technology Studies, Leiden University, The Netherlands
{ecknjpvan, waltmanlr}@cwts.leidenuniv.nl

VOSviewer is a computer program for creating, visualizing, and exploring bibliometric maps of science. In this report, the new text mining functionality of VOSviewer is presented. A number of examples are given of applications in which VOSviewer is used for analyzing large amounts of text data.

## 1. Introduction

VOSviewer is a computer program that we have developed for creating, visualizing, and exploring bibliometric maps of science (Van Eck & Waltman, 2010). The program is freely available at [www.vosviewer.com](www.vosviewer.com). VOSviewer can be used for analyzing all kinds of bibliometric network data, for instance citation relations between publications or journals, collaboration relations between researchers, and co-occurrence relations between scientific terms. In this report, we present the new text mining functionality of VOSviewer. We show how this functionality can be used for analyzing large amounts of text data.

## 2. Text mining functionality

In September 2011, version 1.4.0 of VOSviewer was released. This new version of VOSviewer includes extensive text mining functionality. The text mining functionality of VOSviewer provides support for creating term maps based on a corpus of documents. A term map is a two-dimensional map in which terms are located in such a way that the distance between two terms can be interpreted as an indication of the relatedness of the terms. In general, the smaller the distance between two terms, the stronger the terms are related to each other. The relatedness of terms is determined based on co-occurrences in documents. These documents can be for instance scientific publications (either titles and abstracts or full texts), patents, or newspaper articles. VOSviewer can only handle English language documents.

To create a term map based on a corpus of documents, VOSviewer distinguishes the following steps:

1. Identification of noun phrases. The approach that we take is similar to what is reported in an earlier paper (Van Eck, Waltman, Noyons, & Buter, 2010). We first perform part-of-speech tagging (i.e., identification of verbs, nouns, adjectives, etc.). The Apache OpenNLP toolkit (http://incubator.apache.org/opennlp/) is used for this purpose. We then use a linguistic filter to identify noun phrases. Our filter selects all word sequences that consist exclusively of nouns and adjectives and that end with a noun (e.g., *paper*, *visualization*, *interesting result*, and *text mining*, but not *degrees of freedom* and *highly cited publication*). Finally, we convert plural noun phrases into singular ones.
2. Selection of the most relevant noun phrases. The selected noun phrases are referred to as terms. We have developed a new technique for selecting the most relevant noun phrases. The essence of this technique is as follows. For each noun phrase, the distribution of (second-order) co-occurrences over all noun phrases is determined.

This distribution is compared with the overall distribution of co-occurrences over noun phrases. The larger the difference between the two distributions (measured using the Kullback-Leibler distance), the higher the relevance of a noun phrase. Intuitively, the idea is that noun phrases with a low relevance (or noun phrases with a general meaning), such as *paper*, *interesting result*, and *new method*, have a more or less equal distribution of their (second-order) co-occurrences. On the other hand, noun phrases with a high relevance (or noun phrases with a specific meaning), such as *visualization*, *text mining*, and *natural language processing*, have a distribution of their (second-order) co-occurrences that is significantly biased towards certain other noun phrases. Hence, it is assumed that in a co-occurrence network noun phrases with a high relevance are grouped together into clusters. Each cluster may be seen as a topic.
3. Mapping and clustering of the terms. We use our unified framework for mapping and clustering in this step (Van Eck, Waltman, Dekker, & Van den Berg, 2010; Waltman, Van Eck, & Noyons, 2010).
4. Visualization of the mapping and clustering results. VOSviewer offers various types of visualizations. The program has zoom, scroll, and search functionality to support a detailed examination of a term map.

## 3. Applications

To illustrate the text mining functionality of VOSviewer, we present three examples of applications of this functionality. The term maps that we discuss can be explored in more detail online at www.vosviewer.com/maps/term_maps/.

In the first example, a term map was created based on a corpus of scientific publications in the field of library and information science (LIS). The corpus was extracted from the Web of Science database and consists of the titles and abstracts of about 10,000 publications that appeared in the period 1999–2008 (for more details, see Waltman et al., 2010). Out of the 2101 noun phrases that occur in at least 15 publications in the corpus, the term map contains the 1000 noun phrases that are considered most relevant.

The term map is shown in Figure 1. Colors indicate the density of terms, ranging from blue (lowest density) to red (highest density). As can be seen in Figure 1, examples of prominent terms in LIS research include *journal*, *science*, and *citation*, (upper left), *librarian* and *student* (lower left), and *document*, *task*, and *query* (middle right). These are all single-word terms. Among the slightly less prominent terms, we also observe various multi-word ones, such as *impact factor* (upper left), *information literacy* (lower left), and *search engine* and *test collection* (middle right). The term map also reveals a clear structure of the field. There are three well-separated subfields, which may be referred to as bibliometrics/scientometrics (upper left), library science (lower left), and information science/information retrieval (middle right). The subfields are roughly of equal size. The connection between the bibliometrics subfield and the library science subfield appears to be slightly stronger than the connection of either of these subfields with the information science subfield.

The same term map is also shown in Figure 2. This time colors indicate the research activities of Leiden University. Colors range from blue to red. A blue term is a term that occurs in no or almost no publications of Leiden University in the field of LIS. A red term is a term that occurs relatively frequently in publications of Leiden University. As expected, the research activities of Leiden University turn out to be strongly focused on the bibliometrics subfield.

A second example of an application of the text mining functionality of VOSviewer is shown in Figure 3. The term map shown in this figure was created based on the full text of a PhD thesis on bibliometric mapping of science (Van Eck, 2011). Each paragraph of the full text was treated as a separate document. The 218 most relevant noun phrases were included in the term map. The colors of the 218 terms indicate clusters of related terms identified by VOSviewer. The clusters turn out to correspond reasonably well with the different chapters of the thesis. For instance, the blue cluster (upper left) represents a chapter on automatic term identification and the orange cluster (middle right) represents a chapter on VOSviewer.

Finally, we consider an application in which the text mining functionality of VOSviewer is used to get some insight into the citation impact of the different topics covered by a journal. The journal that we consider is the *Journal of the American Society for Information Science and Technology* (*JASIST*). The analysis uses data from the Scopus database. Using the text mining functionality of VOSviewer, a term map was created based on the titles and abstracts of all publications that appeared in *JASIST* in the period 2005–2009. The term map contains 468 terms and is shown in Figure 4. The color of a term indicates the average citation impact of the publications in which the term occurs. Colors range from blue (lowest citation impact) to red (highest citation impact). Interestingly, there turn out to be large differences in citation impact among the various topics covered by *JASIST*. The term map indicates a strong separation between bibliometric/scientometric topics on the one hand and information science/information retrieval topics on the other hand. On average, publications on bibliometric/scientometric topics turn out to receive many more citations than publications on information science/information retrieval topics.

## 4. Conclusion

In this report, the new text mining functionality of VOSviewer has been presented. A number of examples have been given of applications in which VOSviewer is used for analyzing large amounts of text data. The examples have focused on scientific texts, but of course the text mining functionality of VOSviewer can also be applied to all kinds of non-scientific texts (e.g., newspaper articles).

We hope that the text mining functionality of VOSviewer will be useful to the bibliometric and scientometric community. We very much welcome feedback from users of our software.

## References

Van Eck, N.J. (2011). *Methodological advances in bibliometric mapping of science*. PhD thesis, Erasmus University Rotterdam.
Van Eck, N.J., & Waltman, L. (2010). Software survey: VOSviewer, a computer program for bibliometric mapping. *Scientometrics*, *84*(2), 523–538.
Van Eck, N.J., Waltman, L., Dekker, R., & Van den Berg, J. (2010). A comparison of two techniques for bibliometric mapping: Multidimensional scaling and VOS. *Journal of the American Society for Information Science and Technology*, *61*(12), 2405–2416.
Van Eck, N.J., Waltman, L., Noyons, E.C.M., & Buter, R.K. (2010). Automatic term identification for bibliometric mapping. *Scientometrics*, *82*(3), 581–596.
Waltman, L., Van Eck, N.J., & Noyons, E.C.M. (2010). A unified approach to mapping and clustering of bibliometric networks. *Journal of Informetrics*, *4*(4), 629–635.

Figure 1. Term map of the field of library and information science. Colors indicate the density of terms.

Figure 2. Term map of the field of library and information science. Colors indicate the research activities of Leiden University.

Figure 3. Term map of the full text of Van Eck (2011). Colors indicate clusters of related terms.

Figure 4. Term map of the *Journal of the American Society for Information Science and Technology*. The color of a term indicates the average citation impact of the publications in which the term occurs.